\documentclass{aastex62}
\usepackage{amsmath,amssymb,CJK}

\received{--}
\revised{--}
\accepted{--}
\submitjournal{ApJL}

\shorttitle{The 2013 Outburst of 289P/Blanpain}
\shortauthors{Ye \& Clark}
\begin{document}
\begin{CJK*}{UTF8}{gbsn}

\title{Rising from Ashes or Dying Flash? Mega Outburst of Small Comet 289P/Blanpain in 2013\footnote{Data and codes that generate the figures of this work are available on\dataset[105281/zenodo.3238650]{105281/zenodo.3238650} and \url{https://github.com/Yeqzids/blanpain-2013}.}}

\correspondingauthor{Quanzhi Ye}
\email{qye@caltech.edu}

\author[0000-0002-4838-7676]{Quanzhi Ye (叶泉志)}
\affiliation{Division of Physics, Mathematics and Astronomy, California Institute of Technology, Pasadena, CA 91125, U.S.A.}
\affiliation{Infrared Processing and Analysis Center, California Institute of Technology, Pasadena, CA 91125, U.S.A.}

\author[0000-0002-1203-764X]{David L. Clark}
\affiliation{Department of Earth Sciences, University of Western Ontario, London, Ontario, N6A 5B7, Canada}
\affiliation{Department of Physics and Astronomy, University of Western Ontario, London, Ontario, N6A 3K7, Canada}
\affiliation{Centre for Planetary Science and Exploration, University of Western Ontario, London, Ontario, N6A 5B7, Canada}


\begin{abstract}
Jupiter-family comet 289P/Blanpain was first discovered in 1819 and was then lost for $\sim200$ years, only to be rediscovered in 2003 as a small, weakly active comet. The comet is associated with the Phoenicids, an otherwise minor meteor shower that produced significant outbursts in 1956 and 2014. The shower points to the existence of significant mass-loss events of P/Blanpain in recent history. P/Blanpain was recovered during an apparent large outburst in 2013 July at an appreciable heliocentric distance of 3.9~au, with brightness increase of 9~mag, making it one of the largest comet outbursts ever observed. Here we present an analysis of archival data taken by several telescopes. We find that the 2013 outburst has produced $\sim10^{8}$~kg of dust, which accounts for a modest fraction ($\sim1\%$) of the mass of P/Blanpain's nucleus as measured in 2004. Based on analysis of long-term lightcurve and modeling of coma morphology, we conclude that the 2013 outburst was most likely driven by the crystallization of amorphous water ice triggered by a spin-up disruption of the nucleus. Dust dynamical model shows that a small fraction of the dust ejecta will reach the Earth in 2036 and 2041, but are only expected to produce minor enhancements to the Phoenicid meteor shower. The 2013 outburst of P/Blanpain, though remarkable for a comet at small sizes, does not necessary imply a catastrophic disruption of the nucleus. The upcoming close encounter of P/Blanpain in 2020 January will provide an opportunity to examine the current state of the comet.
\end{abstract}



\section{Introduction} \label{sec:intro}

The common end states for both active and dormant comets include dynamical ejection from the Solar System, solar/planetary impact, and physical disruption. For short-period comets, physical disruptions are several orders of magnitude more frequent than dynamical ejections or impacts \citep{Jewitt2004c}. Well-known examples include 3D/Biela \citep{Jenniskens2007, Wiegert2013}, 73P/Schwassmann-Wachmann 3 \citep{Wiegert2005, Vaubaillon2010}, and 332P/Ikeya-Murakami \citep{Jewitt2016, Kleyna2016, Hui2017}.

289P/Blanpain was first discovered by Jean-Jacques Blanpain in 1819 as a bright comet. It was last observed on 1820 January 15, and was then subsequently lost for nearly 200 years. In 2003, the Catalina Sky Survey discovered 2003 WY$_{25}$, a small asteroid whose orbit closely resembles the orbit of P/Blanpain \citep{Ticha2003, Foglia2005}. The orbits of both the 1819 object and 2003 WY$_{25}$ also match the orbit of the Phoenicid meteor shower. Without contrary evidence, it is normally assumed that 2003 WY$_{25}$ is the remnant of the original P/Blanpain which disrupted in 1819 and supplied the Phoenicids \citep{Jenniskens2005, Watanabe2005, Fujiwara2017}.

Observations of P/Blanpain collected during the 2003/04 apparition showed that it was 1/5 the size of the 1819 object \citep{Jenniskens2005}. The object appeared point-like in nearly all observations, except for the deep integration obtained by \citet{Jewitt2006} in 2004 March. They noted ``a weak optical coma'' and derived a mass loss rate of $10^{-2}~\mathrm{kg~s^{-1}}$, among the lowest values of known comets. With a diameter of $\sim320$~m, the object is also among the smallest cometary nuclei ever observed.

The cometary nature of P/Blanpain became more conclusive when it was recovered by the Panoramic Survey Telescope and Rapid Response System (Pan-STARRS) survey on 2013 July 4, with a distinct coma and a broad tail \citep{Williams2013}. The Pan-STARRS team initially reported a brightness of $V=20.1$ but later commented that it was underestimated (R. Weryk, private communication). Follow-up observations showed an evolving coma and tail, with a brightness at about $V=17.5$ on UT 2013 July 6.55 (H. Sato, iTelescope at Siding Spring). No further observations were made after 2013 July 17, likely because the comet had faded. The comet was near the opposition at that time and should have been easily detectable.

Intriguingly, P/Blanpain was at 3.9~au from the Sun at the time of the Pan-STARRS recovery, with an expected magnitude $V=26.9$\footnote{Calculated using a simple $HG$ model \citep[c.f.][]{Li2015} taking $H_R=21.2$ \citep{Jewitt2006}, assuming $G=0.15$ and a solar color.}. It is therefore evident that P/Blanpain was recovered during a large outburst. With a brightness increase of $\Delta m\approx9$~mag, this is one of the largest cometary outbursts ever observed, exceeded only by the outburst of 17P/Holmes in 2007 ($\Delta m=15$).

\section{Observations} \label{sec:obs}

We identified images containing P/Blanpain obtained during and after its 2013 outburst, using the Solar System Object Image Search provided by the Canadian Astronomy Data Centre \citep{Gwyn2012}. A total of five nights of data, taken by the Canada-France-Hawaii Telescope (CFHT) MegaCam imager and the Dark Energy Camera (DECam) and spanning from 2013 to 2015, were found. For nights with multiple images, we stacked the images following the motion of the comet to enhance the signal of the comet. Brightnesses of P/Blanpain (or, for the case of non-detection, upper limit of the brightnesses) were calculated using the photometric zero-points and color corrections supplied with each image. Details are tabulated in Table~\ref{tbl:obs}.

\begin{deluxetable*}{llcccccccc}
\tablecaption{Summary of the archival observations. \label{tbl:obs}}
\tablehead{
\colhead{Date (UT)} & \colhead{Telescope} & \colhead{Filter} & \colhead{$N$} & \colhead{$r_\mathrm{H}$ (au)} & \colhead{$\varDelta$ (au)} & \colhead{$\alpha$} & \colhead{$m_\mathrm{obs}$} & \colhead{$m_\mathrm{model}$} & \colhead{Detection?}
}
\startdata
2013 July 5 & CFHT & $r_\mathrm{S}$ & 5 & 3.881 & 2.878 & $3^\circ$ & $17.61\pm0.01~r \Leftrightarrow 17.8~V $ & $26.9~V$ & \checkmark \\
2015 January 10 & DECam & $z$ & 1 & 1.956 & 1.933 & $29^\circ$ & $>22.4~z \Leftrightarrow 22.4~V $ & $25.5~V$ & $\times$ \\
2015 February 2 & DECam & \underline{$VR$} & 2 & 2.158 & 1.830 & $27^\circ$ & $>23.2~V$ & $25.5~V$ & $\times$ \\
2015 July 14 & CFHT & $r$ & 1 & 3.342 & 3.206 & $18^\circ$ & $>24.8~r \Leftrightarrow 25.0~V$ & $27.4~V$ & $\times$ \\
2015 July 20 & CFHT & $r$ & 1 & 3.379 & 3.330 & $17^\circ$ & $>24.7~r \Leftrightarrow 24.9~V$ & $27.4~V$ & $\times$ \\
\enddata
\tablecomments{Listed are date, telescope and filter used for the observation, number of images ($N$), heliocentric and geocentric distances ($r_\mathrm{h}$, $\varDelta$) as well as phase angles ($\alpha$) at the time of the observation, observed total brightness or $3\sigma$ upper limit of the comet ($m_\mathrm{obs}$), the brightness of P/Blanpain predicted by a simple $HG$ model taking $H_R=21.2$ \citep{Jewitt2006} and assuming $G=0.15$ ($m_\mathrm{model}$), and whether P/Blanpain is visible. Magnitudes are converted to Johnson $V$ using the relations derived by \citet{Tonry2012} assuming a solar color \citep{Willmer2018}. For the \underline{$VR$}-band observations, we assume their color coefficients equal to those of $V$.}
\end{deluxetable*}

\begin{figure}
\includegraphics[width=0.4\textwidth]{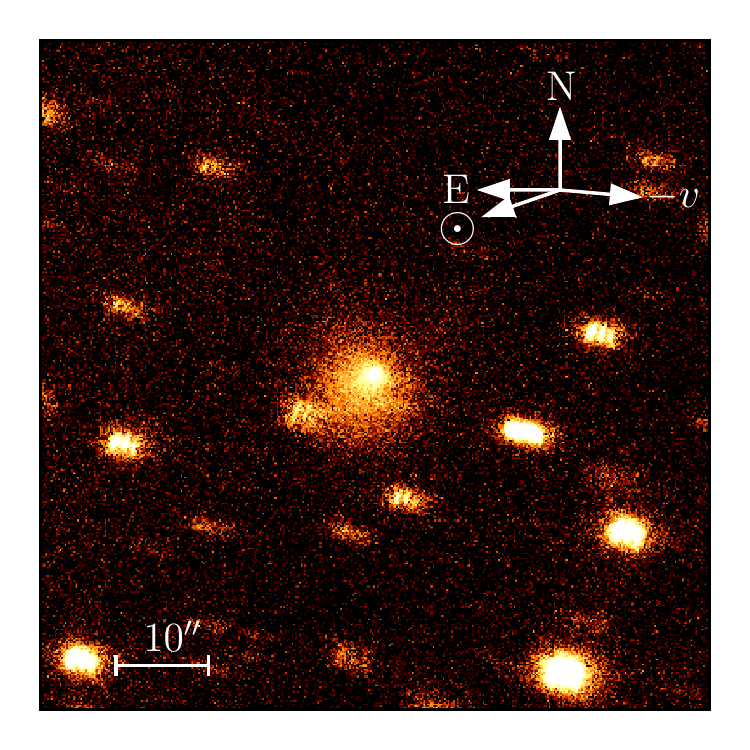}
\caption{Stacked CFHT image of P/Blanpain taken on 2013 July 5 (center). The arrows on the upper-right corner mark the celestial North, celestial East, direction to the Sun ($\odot$), and the minus heliocentric velocity motion ($-v$). Notebook is available \href{https://github.com/Yeqzids/blanpain-2013/blob/master/cfht_stack.ipynb}{here}. \label{fig:cfht}}
\end{figure}

The only set of images where P/Blanpain is clearly visible is the CFHT image set taken on 2013 July 5. The final stacked image is shown as Figure~\ref{fig:cfht}. The coma measures approximately $15-20''$ in diameter, and is slightly shifted towards the sunward direction. By using the photometric constants provided along with the images and adopting a $20''$ diameter aperture centered on the centroid of the coma, we obtain $r_\mathrm{SDSS}=17.7\pm0.1$, equivalent to Johnson $V=17.9$ assuming a solar color. The uncertainty is estimated considering the interference from the background star at 8 o'clock direction of the comet. Other sets of images show no trace of P/Blanpain down to the noise floor within the positional uncertainties (which are all at the order of $1''$). From these non-detections, it can be derived that the diameter of the nucleus of P/Blanpain is no larger than 800~m assuming a geometric albedo of 0.04, consistent with the number directly measured by \citet{Jewitt2006}, which is 320~m.

\section{Onset of the Outburst} \label{sec:onset}

To better constrain the onset time of the outburst, we searched the image catalogs of a variety of surveys including Pan-STARRS, CFHT, the Catalina Sky Survey, and all contributions to the Minor Planet Center's Sky Coverage Database, over the time window of 2013 May 1 to September 30, using the image search facility of the  Fireball Retrieval on Survey Telescopic Image (FROSTI) software \citep{Clark2014}. Our $10,000,000+$ image borehole database contains $116,984$ images for the above time window, of which only a small number of Pan-STARRS exposure sets potentially contain the comet (in addition to the CFHT images mentioned previously).  Table~\ref{tbl:frosti} list these exposure sets. Upon our request, Robert Weryk from the Pan-STARRS team kindly examined the images and provided comments on the visibility of the comet, which are also tabulated in Table~\ref{tbl:frosti}. Based on the result of the search as well as Weryk's comments, we conclude that P/Blanpain was already in outburst at least 1 day before the official rediscovery on 2013 July 4, but the 49-day gap between the 2013 July 3 detection and the last image that covers the predicted position of P/Blanpain (2013 May 15) makes it difficult to pinpoint the exact onset time of the outburst. The non-detections on 2013 May 15, August 24 and September 9 sets an upper limit of $V\approx22.5$ of the comet, based on the typical survey depth of Pan-STARRS \citep{Denneau2013}.

\begin{table*}
\begin{center}
\caption{Results of the FROSTI search of possible detections of P/Blanpain between 2013 May 1 to September 30. The comments are quoted from Robert Weryk through private communication (with bracketed clarifications). \label{tbl:frosti}}
\begin{tabular}{lll}
\tableline
Date (UT) & Survey & Comment \\
\tableline
2013 May 15 & Pan-STARRS & Object not visible \\
2013 June 18 & Pan-STARRS & Object in a chip gap \\
2013 July 3 & Pan-STARRS & Object visible but was not initially picked up by Pan-STARRS software \\
2013 July 4 & Pan-STARRS & Date of rediscovery \\
2013 July 9 & Pan-STARRS & Object visible but was not initially picked up by Pan-STARRS software \\
2013 July 17 & Pan-STARRS & Detector issue \\
2013 July 19 & Pan-STARRS & Detector issue \\
2013 August 24 & Pan-STARRS & Object not visible \\
2013 September 9 & Pan-STARRS & Object not visible \\
\tableline
\end{tabular}
\end{center}
\end{table*}

\section{Coma Morphology and Properties} \label{sec:char}

To understand the driving mechanism of the outburst, we first need to probe the properties of the coma. We use the dust dynamics code originally developed by \citet{Ye2016} to model the coma morphology. To probe different ejection mechanisms, we test two ejection models: the classic \citet{Whipple1951} model and the gravitational escape model.

The Whipple model is devised from the assumption that gas drag from water ice sublimation lifts dust from the sunward-side of the nucleus, a process that happens on most comets. However, we note that the result produced by the Whipple model is also numerically compatible with the ejection caused by amorphous-crystalline transition of water-ice \citep{Prialnik2004}. Therefore, this model can be used to describe the dust ejected by either regime.

Under the Whipple model, the speed of the ejected dust follows the relation

\begin{equation}
v_\mathrm{ej} = 0.8 r_\mathrm{H}^{-9/8} \left( \frac{R_\mathrm{N}}{\rho_\mathrm{d} r_\mathrm{d}} \right)^{1/2}
\end{equation}

\noindent where $r_\mathrm{H}$ is the heliocentric distance, $R_\mathrm{N}$ is the radius of the cometary nucleus, and $\rho_\mathrm{d}$, $r_\mathrm{d}$ are the bulk density and radius of the dust, respectively. The inputs are all in SI units except $r_\mathrm{H}$, which is in au. In our simulation, we take $R_\mathrm{N}=160$~m as measured by \citet{Jewitt2006} and assume $\rho_\mathrm{d}=2000~\mathrm{kg/m^3}$ \citep{Rotundi2015}.

Previous studies \citep[c.f.][Fig. 18]{Jewitt2015} suggested that impulsive ejections tend to have constant ejection speeds, therefore we assume in our gravitational escape model that all particles are ejected isotropically at gravitational escape speed which is defined by

\begin{equation}
v_\mathrm{esc} = \left( \frac{2GM_\mathrm{N}}{R_\mathrm{N}} \right)^{1/2}
\end{equation}

\noindent where $G$ is the gravitational constant and $M_\mathrm{N}$ and $R_\mathrm{N}$ are the mass and radius of the cometary nucleus respectively. For P/Blanpain, we derive $v_\mathrm{esc} \approx 0.1$~m/s. We also simplistically assume that for both models, the dust size $r_\mathrm{d}$ follows a simple power-law, with $\mathrm{d} N(r_\mathrm{d})/\mathrm{d} r_\mathrm{d} \propto r_\mathrm{d}^{-3.6}$ \citep[c.f.][]{Fulle2004}, and the dust size ranges from $10~\micron$ to 0.1~m following the results of in-situ measurements of other comets \citep[e.g.][]{Rotundi2015}.

\begin{figure*}
\includegraphics[width=\textwidth]{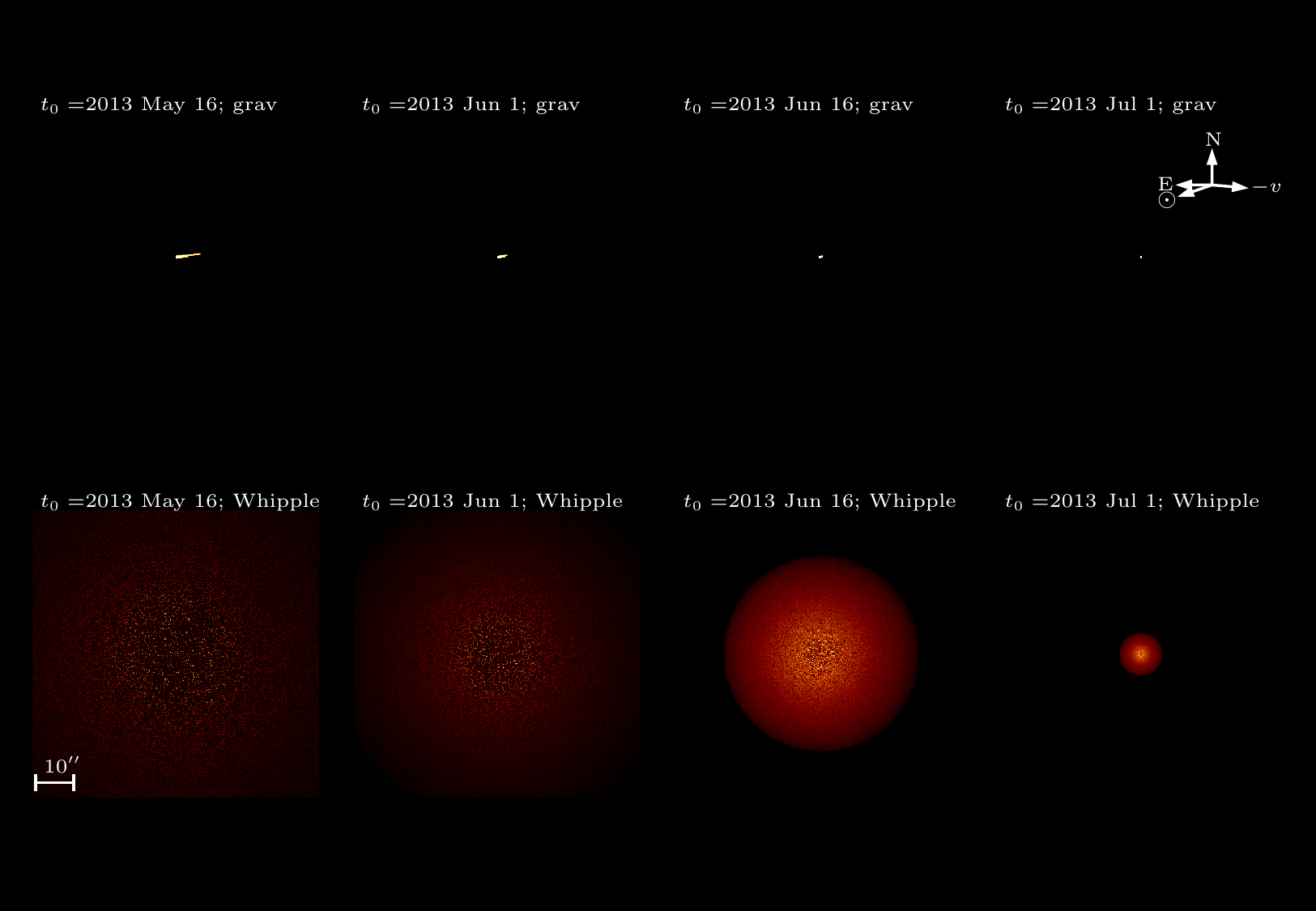}
\caption{Model images of P/Blanpain on 2013 July 5 under different assumptions: ejection dates ($t_0$) of 2013 May 16, June 1, June 16 and July 1, as well as ejections following Whipple model or gravitational escape model. The model most compatible with the observation is the Whipple model with ejection date of 2013 July 1 (lower-right). Notebook is available \href{https://github.com/Yeqzids/blanpain-2013/blob/master/coma_sim.ipynb}{here}. \label{fig:coma-sim}}
\end{figure*}

We test four different dates of which particles are impulsively ejected from the nucleus: 2013 May 16 (i.e. the last pre-outburst observation from Pan-STARRS), June 1, June 16, and July 1, with the model images for each onset date and ejection model shown in Figure~\ref{fig:coma-sim}. The model most compatible with the observation shown in Figure~\ref{fig:cfht} is the Whipple model with ejection date of 2013 July 1, suggesting that the dust were being launched by a sublimation/crystallization-driven activity just a few days before the comet was rediscovered.

While the best-match model reproduces the size and the general shape of the observed coma, we note that it does not reproduce the asymmetry of the coma. This cannot be due to the sunward-only ejection, as this is already accounted for in the assumption of the Whipple model. We suspect that the asymmetry may be due to ejections within a much narrower cone angle near the sub-solar point of the nucleus.

Except for large and very active comets, the brightness of the comet is dominated by the reflected light from the nucleus and the emitted dust. The total mass loss can be calculated by

\begin{equation}
M_\mathrm{d} = \frac{4}{3} \rho_\mathrm{d} \bar{a} C_\mathrm{e}
\end{equation}

\noindent where $\bar{a}=10~\micron$ is the characteristic grain size \citep[c.f.][]{Jewitt2006}, and $C_\mathrm{e}$ is the effective scattering cross-section of the ejecta:

\begin{equation}
\label{eq:cs}
C_\mathrm{e} = \frac{\pi r_\mathrm{H}^2 \varDelta^2}{p_V \Phi(\alpha) a_\mathrm{\oplus}^2} \left( 10^{0.4\Delta m_V} -1 \right) 10^{0.4(m_{\mathrm{\odot, }V} - m_V)}
\end{equation}

\noindent where $p_V=0.04$ is the assumed geometric albedo of the dust, $\Phi{(\alpha)}=0.035\alpha$ is the simple phase function of the target with a phase angle of $\alpha$ \citep[c.f.][]{Li2015}, $a_\mathrm{\oplus}=1.5\times10^{11}$~m is the mean heliocentric distance of the Earth, $\Delta m_V=9$ is the brightness excess in $V$ (as discussed in \S~\ref{sec:intro}), $m_{\mathrm{\odot, }V}=-26.8$ is the apparent $V$ magnitude of the Sun \citep{Willmer2018}, and $m_V$ is the nuclear brightness of P/Blanpain. By substituting corresponding numbers, we obtain $M_\mathrm{d}=1\times10^8$~kg. The uncertainty in this estimate is within a factor of several, mainly contributed by the uncertainty in $\bar{a}$. The mass loss accounts for $\sim1\%$ of the mass of the pre-outburst nucleus ($\sim9\times10^{9}$~kg assuming a nuclear density of $500~\mathrm{kg~m^{-3}}$).

\section{Meteor Prospects} \label{sec:meteor}

The disintegration of comets whose orbits pass near the Earth's orbit is a source of meteor activities at the Earth \citep{Jenniskens2008, Ye2018}. P/Blanpain is associated with the Phoenicid meteor shower \citep{Jenniskens2005, Sato2010, Fujiwara2017}, and it has been shown that the 1819/20 breakup event has produced a short but intense meteor outburst in 1956 \citep{Weiss1958, McBeath2003, Watanabe2005}.

To investigate future encounters between the Earth and the 2013 ejecta, we simulated the dynamical evolution of the ejecta following the same numerical procedure in \citet{Ye2016}. In brief, we simulated a set of particles between $10~\micron$ and 10~cm in diameter, using the RADAU numerical integrator \citep{Everhart1985}, and followed their positions until 2300 January 1. This size range is chosen as it corresponds to size of meteors detectable by modern techniques \citep{Ye2016a}. Ejection vectors of the particles are generated following the Whipple model assuming an ejection time of 2013 July 1, as found in \S~\ref{sec:char}, based on the orbit solution \#7 from the JPL Small-Body Database (\url{https://ssd.jpl.nasa.gov/sbdb.cgi}). Effects considered are the radiation pressure on the particles, as well as the gravitational influences from the Sun and the eight major planets, with the Earth-Moon system represented as a single perturber at the barycenter of the two bodies.

We have identified two encounters between the Earth and the ejecta, as tabulated in Table~\ref{tbl:meteor}. Neither encounters are expected to produce particularly strong meteor activities, as only a small fraction of dust will reach Earth's vicinity. Meteors in both encounters will be dominated by dust of $10~\micron$ in sizes and can only be detected by certain radio techniques (e.g. head-echos).

\begin{table}
\begin{center}
\caption{Predicted encounters of the 2013 ejecta until 2100. \label{tbl:meteor}}
\begin{tabular}{cccccc}
\tableline
Peak time & Duration & Radiant & Geocentric speed & Peak flux & Note \\
& & (J2000) & ($\mathrm{km~s^{-1}}$) & ($\mathrm{km^{-2}~hr^{-1}}$) & \\
\tableline
2036 November 30, 23:50 UT & 8~hr & $\alpha_\mathrm{g}=3^\circ, \delta_\mathrm{g}=-26^\circ$ & 9.5 & $6\times10^{-3}$ & Faint meteors; ZHR$\approx20$ \\
2041 December 1, 3:08 UT & 1~hr & $\alpha_\mathrm{g}=4^\circ, \delta_\mathrm{g}=-25^\circ$ & 9.6 & $2\times10^{-3}$ & Faint meteors; ZHR$\approx10$ \\
\tableline
\end{tabular}
\end{center}
\tablecomments{ZHR is the equivalent Zenith Hourly Rate calculated from the peak flux assuming a power-law distribution with a size index of $-2.8$ \citep{Koschack1990}.}
\end{table}

\section{Mechanism} \label{sec:mec}

Sub-km comets are more prone to rotational excitation which may ultimately lead to their disruption. Following \citet[][Equation 12]{Jewitt2004c}, the excitation timescale of cometary nuclei is 

\begin{equation}
\tau_\mathrm{ex} = \frac{2 \pi \rho_\mathrm{N} r_\mathrm{N}^4}{P_\mathrm{N} V_\mathrm{th} k_\mathrm{T} \dot{M}}
\end{equation}

\noindent where $\rho_\mathrm{N}=500~\mathrm{kg~m^{-3}}$ is the density of the nucleus \citep{Paetzold2016}, $r_\mathrm{N}=160$~m is the radius of the nucleus \citep{Jewitt2006}, $P_\mathrm{N}=5$~hr is the assumed rotational period of the nucleus, $V_\mathrm{th}=500~\mathrm{m~s^{-1}}$ is the thermal speed of the sublimating gas, $0.005<k_\mathrm{T}<0.04$ is the moment-arm of the torque \citep{Belton2011}, and $\dot{M}=0.01~\mathrm{kg~s^{-1}}$ is the mass loss rate measured in 2004 \citep{Jewitt2006}. We derive $\tau_\mathrm{ex}=20-140$~yr, which is in line with the time elapsed since the last observed disruption of P/Blanpain ($\sim200$~yr ago). We acknowledge that time-domain surveys only began after the 1990s, meaning that any outburst before the 1990s would likely have been missed. However, an outburst rate of once every several decades is still compatible with the derived $\tau_\mathrm{ex}$.

In \S~\ref{sec:char} we showed that the morphology of the coma is best explained by a sublimation and/or crystallization regime. The impulsive nature of the outburst, coupled with a heliocentric distance marginally beyond the water-ice sublimation line, seems to disfavor the sublimation-driven scenario. The molecule production rate of the sublimation of pure water-ice at 3.9~au is $\sim4\times10^{20}~\mathrm{molecule~m^{-2}~s^{-1}}$ \citep{Cowan1979}. Assuming an event duration of a few days (inferred from the rapid fading of the comet, see \S~\ref{sec:intro}) and taking the previously derived dust production of $1\times10^8$~kg (equal to a dust production rate at the order of $\sim0.001~\mathrm{kg~m^{-2}~s^{-1}}$), a sublimation-driven regime will lead to an unrealistic dust-to-ice ratio ($\sim10^3$). The amorphous-crystalline transition of water-ice, on the other hand, has been proposed to explain large-scale cometary outbursts as well as the activity of comets beyond the ice line \citep[c.f.][]{Prialnik2004}. Such a process, probably triggered by a rotational breakup of the nucleus, provides a consistent picture of the 2013 outburst of P/Blanpain.

The amorphous-crystalline transition can also be triggered by thermal shocks induced by the rotation and orbital motion of the comet. By using dimensional analysis, we derive a skin depth of $(\kappa P)^{1/2}\sim4$~m (where $\kappa\sim10^{-7}~\mathrm{m^2~s^{-1}}$ is the thermal diffusivity, and $P=5$~yr is the orbital period of the comet). If P/Blanpain only reached its current orbit very recently, amorphous ice located $\gg4$~m below the surface would have been largely unperturbed by the thermal shocks, and could be the source of the 2013 outburst if located at a suitable depth. Preliminary dynamical simulation shows that P/Blanpain had a higher perihelion ($q\gtrsim2$~au) a few $10^3$~yr ago, consistent with the abovementioned assumption. Given that the characteristic timescale of thermal excitation is $\propto r_\mathrm{N}^2/\kappa$, smaller nuclei should be more prone to rotational disruption (whose timescale $\propto r_\mathrm{N}^4$).

\section{Conclusion}

A 9-magnitude outburst of the small, 0.3-km-diameter comet P/Blanpain at an appreciable heliocentric distance (3.9~au) is one of the largest cometary outbursts ever observed. Despite the magnitude of the outburst, our analysis showed the ejected material only accounts for $\sim1\%$ of the total mass of the nucleus, therefore the nucleus likely has survived the outburst. This echoes a few previous examples of multi-magnitude outbursts exhibited by sub-kilometer-sized comets, of which the comets have seemingly survived \citep{Ye2017}.

The observed coma morphology and lightcurve matches an impulsive ejection of dust likely driven by the crystallization of amorphous water ice. Such a process can be triggered by rotational breakup of the nucleus. Smaller fragments generated from the disruption, if any, could have a rotational excitation timescale of $\ll100$~yr, and may exceed their own critical rotation periods within our lifetime.

We found that the bulk of the material released in the 2013 event will not reach the Earth in the next $\sim300$~yr. A small fraction of the material, dominated by $10~\micron$-sized dust, will encounter the Earth on 2036 December 1 and 2041 December 1, and could produce minor enhancements of the Phoenicid meteor shower. The small sizes of the dust particles, coupled with the low encounter speed, means that the activities will be dominated by very faint meteors best observed by certain radio techniques.

P/Blanpain will have a close encounter with the Earth in 2020 January at a distance of 0.09~au. Preliminary dynamical simulation shows that this is one of P/Blanpain's closest encounters to the Earth, before a close encounter with Jupiter in the year of 2292 that will move the comet to the outer solar system (Ye et al. in prep). Observations during this close approach will likely reveal the current state of P/Blanpain and provide information about cometary breakups.

%\begin{figure}
%\includegraphics[width=0.5\textwidth]{q.pdf}
%\caption{Evolution of perihelion distance $q$ of P/Blanpain using the orbit solution derived in \S~\ref{sec:meteor}. %Filled area is the $1\sigma$ uncertainty of the evolution. \label{fig:q}}
%\end{figure}




\acknowledgments

We thank an anonymous referee for a careful review. We are grateful to Robert Weryk for checking the Pan-STARRS images at our request, and to Davide Farnocchia for updating the JPL orbit solution for P/Blanpain. Q.-Z. Ye is supported by the GROWTH project funded by the National Science Foundation under Grant No. 1545949.

This research used the facilities of the Canadian Astronomy Data Centre operated by the National Research Council of Canada with the support of the Canadian Space Agency.

This research used observations obtained with MegaPrime/MegaCam, a joint project of CFHT and CEA/DAPNIA, at the Canada-France-Hawaii Telescope (CFHT) which is operated by the National Research Council (NRC) of Canada, the Institut National des Science de l'Univers of the Centre National de la Recherche Scientifique (CNRS) of France, and the University of Hawaii.

This research also used data obtained with the Dark Energy Camera (DECam), which was constructed by the Dark Energy Survey (DES) collaborating institutions: Argonne National Lab, University of California Santa Cruz, University of Cambridge, Centro de Investigaciones Energeticas, Medioambientales y Tecnologicas-Madrid, University of Chicago, University College London, DES-Brazil consortium, University of Edinburgh, ETH-Zurich, University of Illinois at Urbana-Champaign, Institut de Ciencies de l'Espai, Institut de Fisica d'Altes Energies, Lawrence Berkeley National Lab, Ludwig-Maximilians Universitat, University of Michigan, National Optical Astronomy Observatory, University of Nottingham, Ohio State University, University of Pennsylvania, University of Portsmouth, SLAC National Lab, Stanford University, University of Sussex, and Texas A\&M University. Funding for DES, including DECam, has been provided by the U.S. Department of Energy, National Science Foundation, Ministry of Education and Science (Spain), Science and Technology Facilities Council (UK), Higher Education Funding Council (England), National Center for Supercomputing Applications, Kavli Institute for Cosmological Physics, Financiadora de Estudos e Projetos, Funda\c{c}\~{a}o Carlos Chagas Filho de Amparo a Pesquisa, Conselho Nacional de Desenvolvimento Cient\'{i}fico e Tecnol\'{o}gico and the Minist\'{e}rio da Ci\^{e}ncia e Tecnologia (Brazil), the German Research Foundation-sponsored cluster of excellence ``Origin and Structure of the Universe'' and the DES collaborating institutions.

This research has made use of data and/or services provided by the International Astronomical Union's Minor Planet Center.

\facilities{Blanco, CFHT}
\software{Astropy \citep{Astropy2018}, FROSTI \citep{Clark2014}, Jupyter Notebooks \citep{Kluyver2016}, Matplotlib \citep{Hunter2007}, MERCURY6 \citep{Chambers1997}}

\end{CJK*}



\end{document}